\documentclass[aps,pra,amsmath,amssymb,twocolumn,showpacs,
nofootinbib % %comment this out to get footnotes at the end, but it doesn't work :(
]{revtex4-1}
\pdfoutput=1
\usepackage{epsfig}
\usepackage{subfigure}
\usepackage{amsmath, amssymb, amsthm, amsfonts}
\usepackage{bm}
\usepackage{braket}
\usepackage[english]{babel}
\usepackage{hyperref}

\def\ket#1{|#1\rangle}
\def\bra#1{\langle#1|}

\def\Per{\mathrm{Per}}

\begin{document}
\title{Experimental Boson Sampling}
\author{Max Tillmann$^{1,2}$}\email{max.tillmann@univie.ac.at}
\author{Borivoje Daki{\'c}$^{1}$}
\author{Ren{\'e} Heilmann$^{3}$}
\author{Stefan Nolte$^{3}$}
\author{Alexander Szameit$^{3}$}
\author{Philip Walther$^{1,2}$}
\affiliation{$^1$Faculty of Physics, University of Vienna, Boltzmanngasse 5, A-1090 Vienna, Austria}
\affiliation{$^2$Institute for Quantum Optics and Quantum Information, Austrian Academy of Sciences, Boltzmanngasse 3, A-1090 Vienna, Austria}
\affiliation{$^3$Institute of Applied Physics, Abbe Center of Photonics, Friedrich-Schiller Universit\"at Jena, Max-Wien-Platz 1, D-07743 Jena, Germany}

\begin{abstract}
Universal quantum computers promise a dramatic speed-up over classical computers but a full-size realization remains challenging. However, intermediate quantum computational models have been proposed that are not universal, but can solve problems that are strongly believed to be classically hard. Aaronson and Arkhipov have shown that interference of single photons in random optical networks can solve the hard problem of sampling the bosonic output distribution which is directly connected to computing matrix permanents. Remarkably, this computation does not require measurement-based interactions or adaptive feed-forward techniques. Here we demonstrate this model of computation using high--quality laser--written integrated quantum networks that were designed to implement random unitary matrix transformations. We experimentally characterize the integrated devices using an in--situ reconstruction method and observe three-photon interference that leads to the boson-sampling output distribution. Our results set a benchmark for quantum computers, that hold the potential of outperforming conventional ones using only a few dozen photons and linear-optical elements.
\end{abstract}

\maketitle

\section{Introduction}
More than a decade ago, Knill, Laflamme and Milburn (KLM) showed in their seminal work~\cite{Knill2001} that scalable photonic quantum computing is possible using only linear optical circuits, single-photon sources and detectors, and measurement-induced effective nonlinearities. The use of ancillary photons and adaptive feed-forward techniques~\cite{Prevedel2007} not only enables the heralding of successful gate operations~\cite{Gasparoni2004,Okamoto2011} but also provides a basis for protocols in which probabilistic two-photon gates are teleported into a quantum circuit with high probability~\cite{Gao2010a}. This has encouraged researchers to pursue the realization of  large-scale photonic quantum computers~\cite{Ladd2010}. Even though there has been impressive theoretical progress~\cite{Yoran2003,Nielsen2004,Browne2005,Ralph2005}, the required number of indistinguishable ancilla photons for a universal optical quantum computer~\cite{Nielsen&Chuang} appears to be very challenging given current photonic quantum technology.

On the other hand, several interesting intermediate models of quantum computation have been recently proposed~\cite{Knill,Aaronson2004a,Jordan2010}. Even though they do not enable universal quantum computation, these models still provide a dramatic computational speed-up for particular tasks. Such quantum computers can be seen in analogy to quantum emulators that are designed for the simulation of one particular physical system~\cite{2012}. In contrast to the proposed KLM scheme, these models need neither entangling gate operations, adaptive measurements, nor ancilla photons, and are thus technically more feasible. The intermediate quantum computation model proposed by Aaronson and Arkhipov~\cite{Aaronson2010} seems to be extremely resource efficient as it utilizes the unique advantages of the photons' mobility and bosonic nature to solve sampling problems that are believed to be classically hard~\cite{Aaronson2011,Rhode2012}.

Here we experimentally demonstrate the boson-sampling computation based on non-interacting and identical bosons. It is strongly believed that there is no known way of efficiently simulating the output of this computation by classical means~\cite{Aaronson2010}. It is realized by using a quantum system composed of $n$ non-interacting bosons that are processed through a network of $m$ physical modes, where $m>n$. The bosonic nature of the photons themselves leads to non-classical interference when the photons propagate through a random network and produce a complex output probability distribution that is hard to sample on a classical computer. The simplest case of the non-classical interference effect was demonstrated for two photons in the seminal experiment of Hong-–Ou-–Mandel (HOM)~\cite{Hong1987} in 1987. When two single photons enter, one from each input mode, a beam splitter with 50$\%$ reflectivity ($50/50$ beam splitter), they will always exit the beam splitter together in one of the two output modes when the indistinguishability and temporal overlap is perfect (Figure 1a). This quantum physical phenomenon has not only become a standard technique for aligning quantum photonic devices, but also provides, together with the measurement process, the main underlying physical mechanism for non-trivial quantum gates in linear optical quantum computing~\cite{Obrien2003}.
\\ 
\indent This tendency of photons to "bunch" is explained by an effect in which two possible outcomes interfere destructively. Because a beam splitter represents a unitary transformation~\cite{Zeilinger1981}, the probability that two photons entering a 50/50 beam splitter in input modes $a$ and $b$ respectively that both will be transmitted cancels with the probability that they will both be reflected. More generally, the probability of finding one photon in output mode $a'$ and the other in mode $b'$ is given by the permanent of the beam splitter matrix $\mathrm{BS}$:
\\
\\
%%%%%%%%%%%%%%%%%%%%%%%%%%%%%%%%%%%%%%%%%%%%%%%%%%%%%%%%%%%%%%%%%%%%%%%%%%%%%
\begin{equation}
P=|\mathrm{Per}(\mathrm{BS})|^2=|\mathrm{Per}\left(
                             \begin{array}{cc}
                               T & i R \\
                               i R & T \\
                             \end{array}
                           \right)
|^2=|T^2-R^2|^2, 
\end{equation}
%%%%%%%%%%%%%%%%%%%%%%%%%%%%%%%%%%%%%%%%%%%%%%%%%%%%%%%%%%%%%%%%%%%%%%%%%%%%%%%
where $T$ and $R$ are beam splitter transmission and reflection coefficients, respectively. In the case of a $50/50$ beam splitter the permanent is obviously zero, therefore the photons bunch into one of the output modes. The same formula holds for the general case of $n$ photons injected into $n$ different modes of an $m \times m$ optical network with a underlying matrix $U$. The probability that one finds these photons in  $n$ different output modes is given by the permanent of the $n\times n$ sub-matrix $\tilde{U}$ of the unitary $U$, $ P=|\mathrm{Per}(\tilde{U})|^2$ (see Methods for details). For complex networks, such as randomly designed networks, computing the permanent of the underlying unitary matrix on a classical computer is conjectured to scale exponentially in time~\cite{Aaronson2010} with respect to the size of the unitary matrix. The same holds for the output distribution. Thus, even for today's most powerful conventional computers, this puts an upper limit on the size of a unitary matrix for which the output distribution can be calculated. In particular, Rhode and Ralph~\cite{Rhode2012} estimated that for a random optical network of $n \approx 20$ photons in $m \approx 400$ modes, sampling the output distribution is already intractable for conventional computers. Obviously, these phyical requirements seem to be feasible in the not-too-distant future given the recent progress in photonic quantum technology~\cite{OBrien2009}. This visualizes the importance of a benchmark computation with a lower number of photons in random optical networks where the classical verification is still possible.

\begin{figure}
\includegraphics[width=0.45\textwidth]{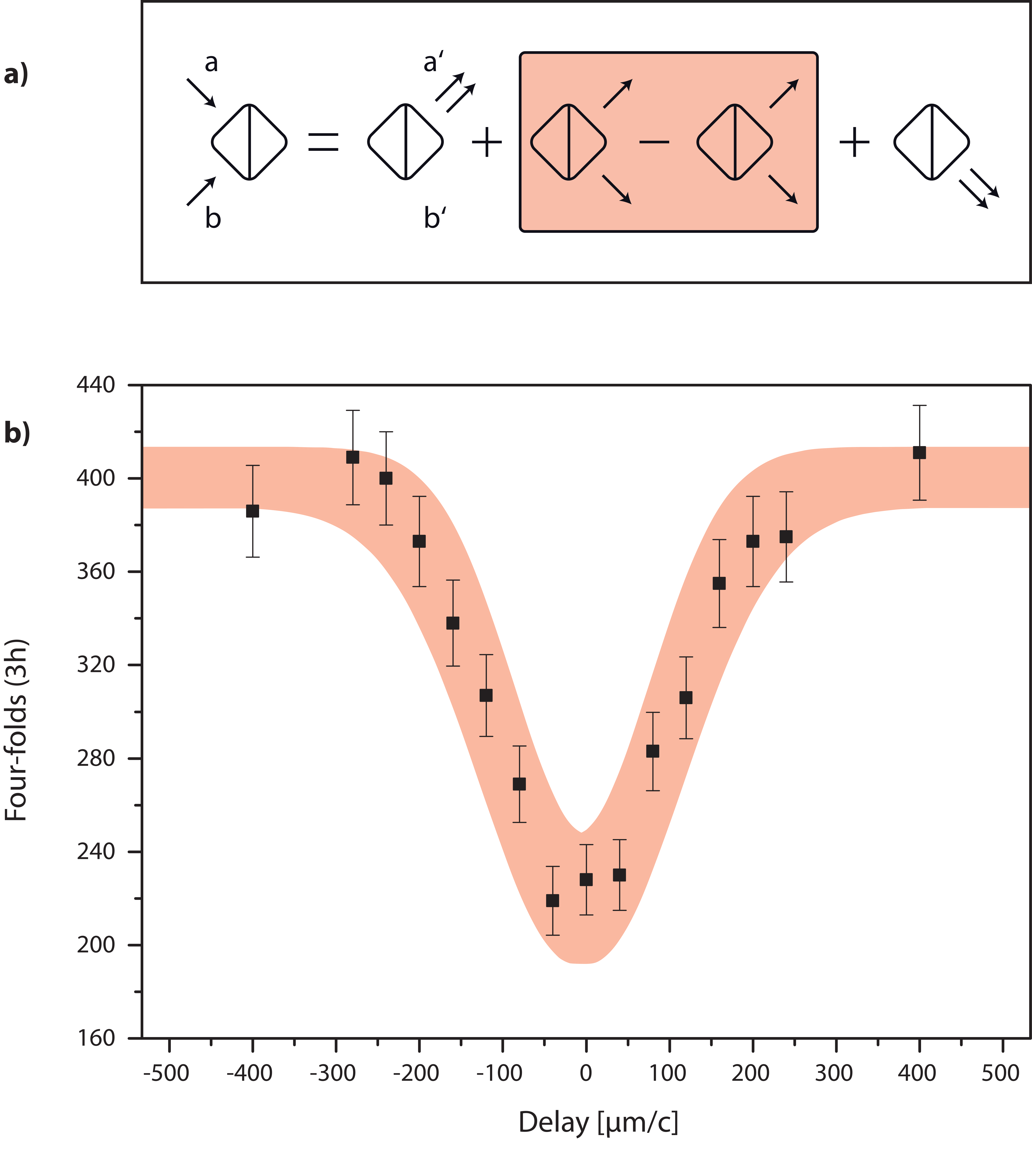}
\caption{\textbf{Non-classical interference.} \textbf{a)} Basic two-photon interference. When two indistinguishable photons enter a $50/50$ beam splitter the two possible outputs, both photons being transmitted or reflected, are interfering destructively (highlighted in red). Therefore the probability of finding two photons in different output modes $a'$ and $b'$ is zero. This is described mathematically as the permanent of the $50/50$ beam splitter unitary. \textbf{b)} Example of an experimental three-photon interference. Three photons were injected into input modes $3$, $4$, and $5$ of one of our optical networks. We measured four-fold coincidence counts between a trigger detector and output modes $2$, $4$, and $5$ of our device, while simultaneously delaying the input photons in mode $4$, and $5$. The Dip shows a clear signature of genuine three-photon interference. The error for the experimental data follows a Poissonian distribution of the measured counts. The shaded area represents the Gaussian fit including errors.}
\label{FIG:Dip}
\end{figure}

Even though bosons tend to bunch, also known as the ``boson birthday paradox''~\cite{Arkhipov2011}, for networks with a sufficiently large number of modes ($m\gg\sqrt{n}$), the probability of detecting $n$ photons in $n$ spatially separated modes as $n$-fold coincidences dominates. This remarkable feature reduces the technological requirements as no number--resolving detectors are needed for this intermediate model of quantum computation and makes a full-fledged boson--sampling computation more feasible in the near future. Therefore, in our experiment we consider only those measurement outcomes where three photons are detected in three spatially separated modes as three-fold coincidences

\section{Experiment}
The boson-sampling computation is demonstrated for different randomly designed optical networks. The computation is initialized by the insertion of three indistinguishable photons, one in each input mode, into an integrated circuit with five input and five output modes. After the propagation through the waveguide structures, the corresponding output distribution is recorded by detecting all possible three-photon coincidence measurements, where each photon is found in a different output mode.

Each integrated circuit was fabricated with a direct laser writing technique~\cite{Itoh2006,Marshall2009} and consists of five spatial modes that are coupled by eight beam splitters and eleven phase shifters (see Methods). To obtain different optical networks, two parameters were randomly varied during the fabrication process: the phases, by adjusting the relative path length between optical elements, and the beam splitter ratios, by tuning the evanescent coupling among the modes. The schematic drawing of these integrated optical networks is shown in Figure 2. The three-photon input state was generated via the process of spontaneous parametric down-conversion~\cite{Kwiat1995}.

The photon source was aligned to emit the entangled state $\ket{\Phi^{+}}=\left(\ket{H}_a \ket{H}_b + \ket{V}_a \ket{V}_b\right)/\sqrt{2}$, where H and V denote horizontal and vertical polarization, respectively, and $a$ and $b$ correspond to the two spatial modes. By pumping with higher power ($700 mW$ cw--equivalent) also two photon pairs are emitted as a four--fold emission, while the even higher--order emission is kept low. To enable a triggered three-photon emission, two photon pairs must be emitted simultaneously into spatial modes $a$ and $b$, resulting in 
\begin{gather*}
\ket{\Psi}_{a,b}=(\ket{HH}_a\ket{HH}_b+\ket{HV}_a\ket{HV}_b\\ +\ket{VV}_a\ket{VV}_b)/\sqrt{3}.
\end{gather*}
These photons are guided to a state preparation stage utilizing two polarizing beam splitters (PBS1 and PBS2) such that a successful detection event in the trigger mode $a''$ heralds the generation of the states $\ket{H}_{a'}\ket{H}_{b'}\ket{V}_{b''}$ or $\ket{VV}_{b'}$. Postselection on a four-fold-coincidence, consisting of the trigger event and three detection events in the output modes of the circuit, ensures that three photons entered the waveguide in separate spatial modes. A half-wave plate in mode $b''$ introducing a $90^{\circ}$ rotation is used to render the photons indistinguishable in polarization. Using mating adapters, the three photons can be inserted in any combination of three input modes of the polarization-maintaining fiber-array that is butt-coupled to the integrated device. A schematic of the experimental setup is shown in Figure 3.

\begin{figure}
\includegraphics[width=0.5\textwidth]{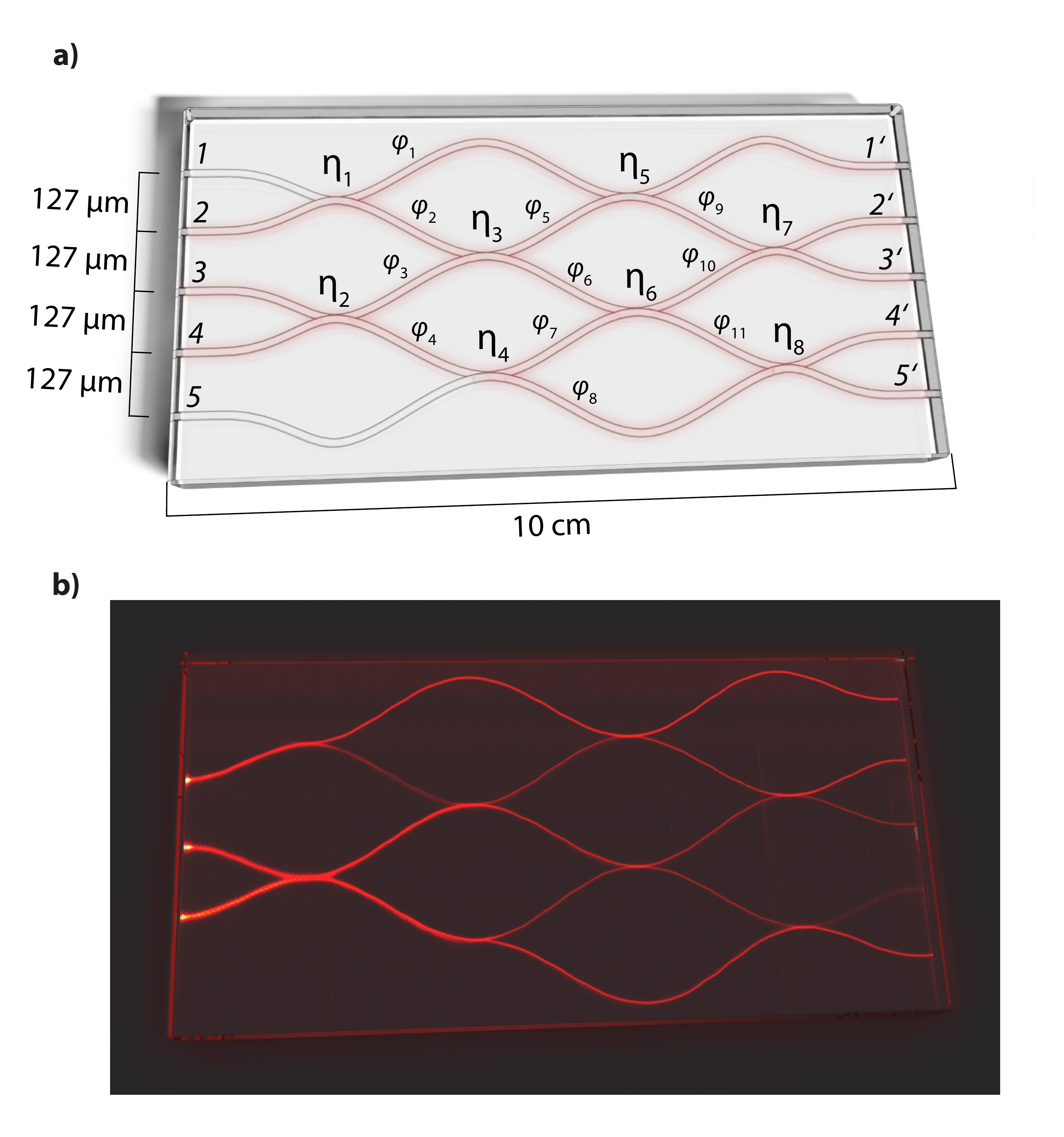}
\caption{\textbf{The optical networks.} a) Schematic drawing: The circuit consists of five input modes ($1$ to $5$), five output modes ($1'$ to $5'$), eight directional couplers ($\eta_1$ to  $\eta_8$) and eleven phase shifters ($\varphi_1$ to $\varphi_{11}$). Up to three single photons can be coherently launched into any combination of input modes. Each output mode is connected to a single-photon detector and coincidences are recorded with a home-built FPGA-logic. Neighboring modes are separated $127 \mu m$ and the chip exhibits a total length of $10 cm$. Three different optical networks written on the same chip were used in the experiment. b) Fluoresence image: In order to visualize the light evolution in the network, coherent laser light at a wavelength of $633nm$ is launched into input modes $2$ to $4$ of an optical network. Color centers are excited by the propagating beam and emit fluorescent light at a wavelength of $650 nm$. The Fluoresence signal is directly proportional to the propagating light intensity.}
\label{FIG:Circuit}
\end{figure}

\section{Result}
The multi-photon interference on chip~\cite{Metcalf2012} is controlled by three adjustable delay lines to temporally overlap the photons. Scanning the temporal delays results in a three-photon HOM-dip that acts as a strong signature of the three photons' non-classical interference (Figure 1 b). The underlying unitary operation of the integrated circuits was reconstructed by using an adaption of a recently proposed method~\cite{Laing2012}. For each optical network, the 19 independent parameters were fitted to the experimentally acquired 25 single-photon probabilities and 40 two-photon visibilities (see Methods for details). Figure 4 depicts the experimental data and theoretical predictions of the boson-sampling computation for two different randomly designed integrated circuits. For each experimental data point, four-fold coincidence events were detected for $20$ hours. The output distributions are in good agreement with the theoretical values obtained from the reconstructed unitary matrices.

\begin{figure*}
\includegraphics[width=1\textwidth]{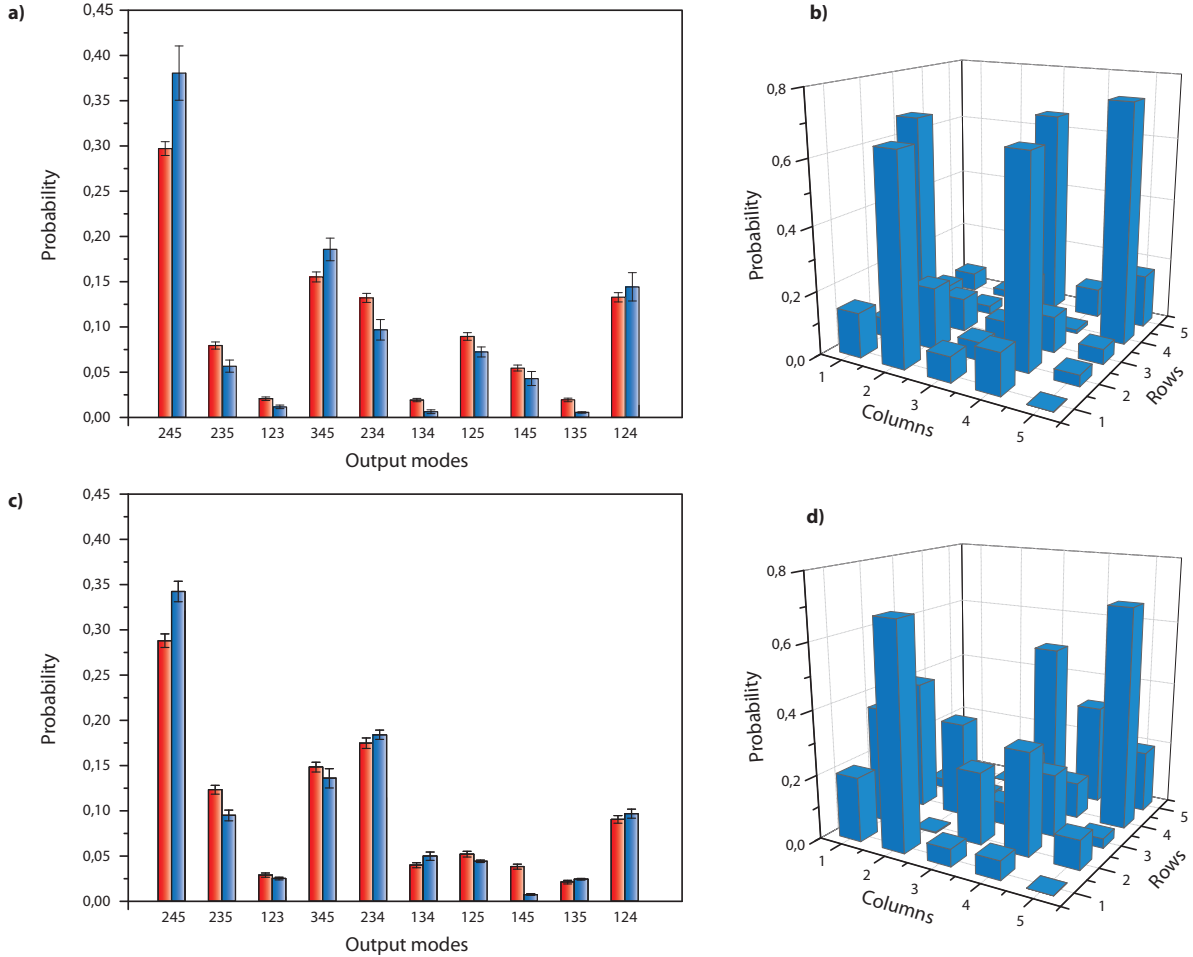}
\caption{\textbf{Three photon probabilities and reconstructed unitary matrices.} Experimentally measured (red on the left) and theoretical (blue on the right) three-photon probabilities for the ten possible output combinations. Photons were injected into input modes $3$, $4$, and $5$ of device no. 1 \textbf{a)} and input modes $3$, $4$, and $5$ of device no. 2 \textbf{c)}. The errors for the probabilities follow a Poissonian distribution. Absolute value squared of the reconstructed matrix elements of \textbf{b)} optical network no.1 and \textbf{d)} optical network no.2.}
\label{FIG:Prob}
\end{figure*}

\section{Conclusion}
Our experiment presents the first benchmark quantum computation on randomly designed optical networks showing a boson-sampling computation. This intermediate model of quantum computation is of particular interest as the bosonic interference of photons in random networks is already hard to simulate on conventional computers. In contrast to universal models of photonic quantum computers that rely on ancilla photons, measurement-induced interactions, and adaptive feed-forward techniques, the boson-sampling computation requires only passive optical elements. This relaxes the physical requirements significantly such that a continuous improvement of current multi-photon sources and detection efficiencies as well as reducing the losses in integrated circuits, might lead to quantum computations in regimes where classical verification is no longer possible in the near future.

\begin{figure*}
\includegraphics[width=0.85\textwidth]{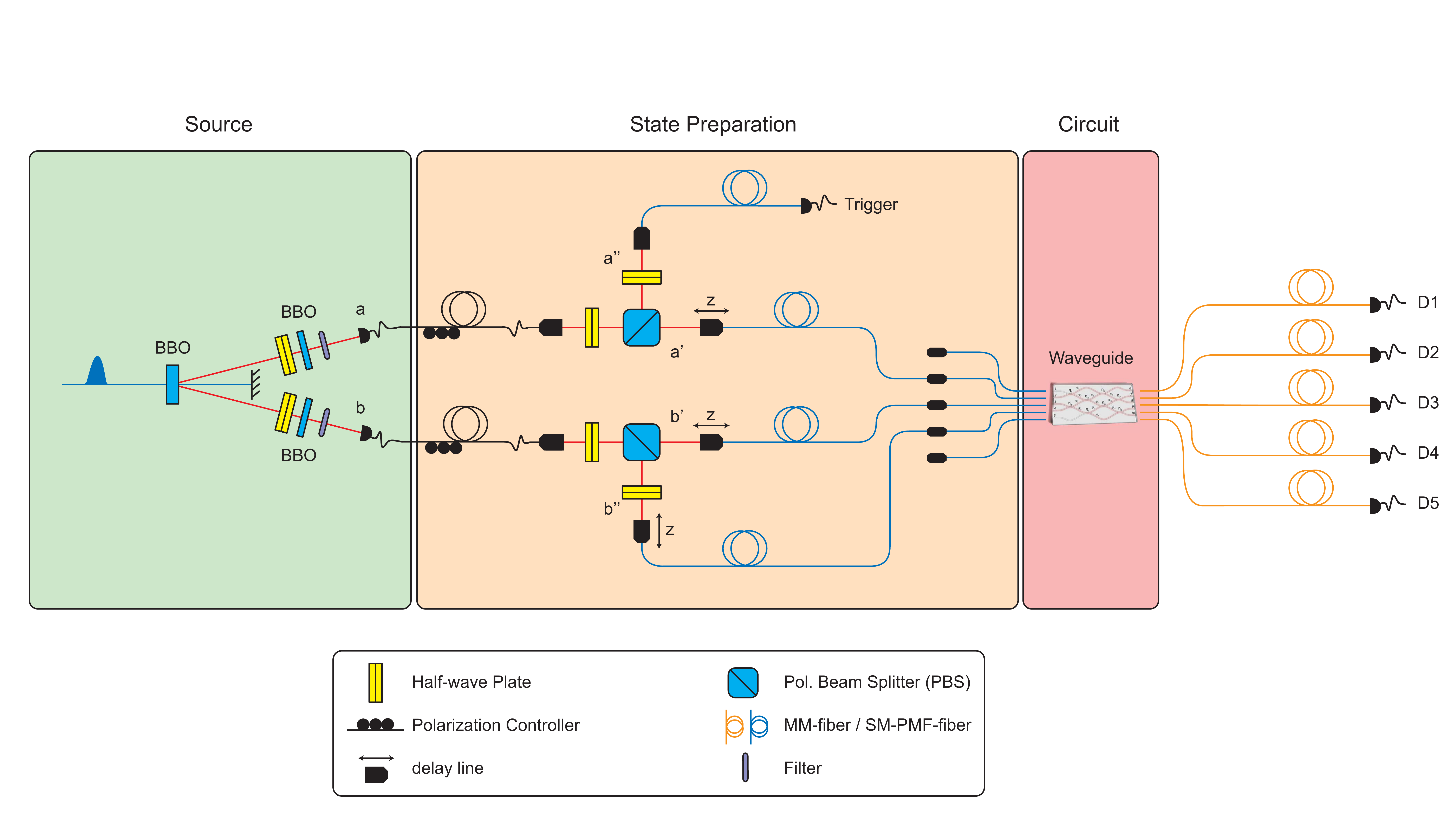}
\caption{\textbf{Experimental setup.} The experimental setup can be divided into three main parts. The first part is a downconversion source pumped with $150 fs$-pulses at a wavelength of $394.5 nm$ and cw-equivalent power of $700 mW$. Four photons are created using higher-order emissions from a $2 mm$ BBO crystal cut for a non-collinear typeII process. Filtering with $\lambda_{FWHM}=3 nm$ filters guarantees spectral indistinguishability. The photons are coupled to single-mode fibers and guided to the state preparation stage. Here, four-fold events are split up via two polarizing beam splitters (PBS) and a half-wave plate (HWP) in mode $b''$ ensures that all photons entering the integrated device exhibit the same polarization. The input photons are coupled into polarization maintaining fibers (PM-fibers), which can be mated to any of the five fibers of the input-fiber array. Temporal overlap in the circuit is achieved with three delay lines. After passing through the waveguide, the photons are coupled to graded-index multimode fibers and sent to single-photon avalanche photo diode detectors (D1 to D5).}
\label{FIG:Setup}
\end{figure*}

\section{Methods}
\textbf{Boson sampling computation.} This intermediate model of quantum computation involves a quantum system of $n$ non-interacting bosons operating between $m$ physical modes, where $m>n$. We define the computation basis states $\ket{i_1,i_2,\dots,i_m}$, where $i_1+i_2+\dots i_m=n$. For example, the state $\ket{2,1,0,1}$ is the state of two bosons in the first mode and one boson in the second and fourth mode. The total number of states in the computational basis $D=\binom{m+n-1}{n}$ is exponentially large in $m$. Since there is no interaction, all the gates in the model are single-particle transformations defined by $m \times m$ complex unitary matrices. For the purposes of this work it is important to recall the definition of the permanent of a $m \times m$ matrix $U$
\begin{equation*}
\Per(U)=\sum_{\sigma\in S_m}\prod_{i=1}^m U_{i,\sigma(i)},
\end{equation*}
where $S_m$ is the set of all permutation of $m$ elements. For example, the permanent of $2\times2$ unitary matrix $U=\left(
                                                                                                              \begin{smallmatrix}
                                                                                                                a & b \\
                                                                                                                c & d \\
                                                                                                              \end{smallmatrix}
                                                                                                            \right)
$ reads $\Per(U)=ad+bc$. If the $m \times m$ matrix $U$ exhibits sufficient complexity, e.g. is not built out of only 0 and 1 elements, than the evaluation of its permanent is strongly believed to be hard on a classical computer. This means that the computation requires an exponential number of steps of computation in $m$. Consider an input state $\ket{I}=\ket{i_1,i_2,\dots,i_m}$ of $n$ bosons in  $m$ modes that is transformed via some unitary $m\times m$ matrix $U$. The probability of finding the state  $\ket{O}=\ket{j_1,j_2,\dots,j_m}$ at the output is given by
\begin{equation*}
P_{I,O}=|\bra{O}U\otimes U\dots\otimes U\ket{I}|^2.
\end{equation*}
Careful analysis~\cite{Aaronson2010} shows that this probability can be expressed trough the matrix permanents:
\begin{equation*}
P_{I,O}=\frac{|\mathrm{Per}(U_{I,O})|^2}{i_1!i_2!\dots i_m!j_1!j_2!\dots j_m!},
\end{equation*}
where the $n\times n$ matrix $U_{I,O}$ is defined in the following manner. First we define the $m\times n$ matrix $U_{I}$ by taking $i_k$ copies of the $k^{\mathrm{th}}$ column of $U$, for each $k=1\dots m$. Then we form the $n\times n$ matrix $U_{I,O}$ by taking $j_k$ copies of the $k^{\mathrm{th}}$ row of $U_{I}$ , for each $k=1\dots m$. As an example, consider a $3\times 3$ matrix
\begin{equation*}
U=\left(
    \begin{array}{ccc}
      a & b & c \\
      d & e & f \\
      g & h & j \\
    \end{array}
  \right),
\end{equation*}
and the input $\ket{I}=\ket{1,1,0}$ and output state $\ket{O}=\ket{0,1,1}$. Than the matrix $U_{I}$ is
\begin{equation*}
U_{I}=\left(
    \begin{array}{cc}
      a & b  \\
      d & e  \\
      g & h  \\
    \end{array}
  \right),
\end{equation*}
and finally $U_{I,O}$ reads
\begin{equation*}
U_{I,O}=\left(
    \begin{array}{cc}
      d & e  \\
      g & h  \\
    \end{array}
  \right).
\end{equation*}
Therefore, the probability of finding the state $\ket{O}=\ket{0,1,1}$ at the output, that is to find one boson in the mode $2$ and one in mode $3$ is given by
\begin{equation*}
P_{I,O}=\frac{|\mathrm{Per}(U_{I,O})|^2}{1!1!0!0!1!1!}=|dh-eg|^2.
\end{equation*}
From the previous example it is clear that for cases where the input and output modes are occupied by at most one boson the matrix $U_{I,O}$ turns out to be the $n\times n$ submatrix of $U$.
\\[\baselineskip]
\textbf{Experimental setup.} An $80 MHz$ Ti:Sapphire oscillator (Chameleon, Coherent Inc.) emitting $150 fs$ pulses at  $789 nm$ ($2.5 W$ cw-equivalent power) is upconverted to $700 mW$ cw-equivalent power at $349.5 nm$ via a $\text{LiB}_3\text{O}_5$ crystal (LBO) (HarmoniXX, A.P.E GmbH). The beam is focused on a $2 mm$ $\beta$-$\text{BaB}_2\text{O}_4$ (BBO) crystal cut for degenerate non-collinear type-II spontaneous parametric downconversion ~\cite{Kwiat1995}. To achieve spectral indistinguishability, the downconverted photons are filtered by interference filters ($\lambda_{FWHM} = 3 nm$) and collected with single mode fibers.
The cw-equivalent pump power of $700 mW$ allows for emission of four-fold states at high count rates. Noise contribution from higher-order terms was measured to be lower than four percent. The double-pair emission of the source generates the following state

\begin{gather*}
 \ket{\Psi}_{a,b}=(\ket{HH}_a\ket{HH}_b + \ket{VV}_a\ket{VV}_b\\
 +\ket{HV}_a\ket{HV}_b)/\sqrt{3}.
\end{gather*}
Two polarizing beam splitters (PBS1 and PBS2) distribute the photons into four modes such that a four-fold coincidence postselection guarantees that three photons entered the waveguide in different modes. Half-wave plates ensure indistinguishability in polarization. Three delay lines are used to temporally overlap the photons for non-classical interference. The photons reach the integrated circuit via a polarization maintaining v-groove fiber array that is butt-coupled to the waveguide. Index matching gel is applied to reduce reflection-losses. A graded-index multimode fiber array is butt-coupled to the output of the waveguide and connected to single-photon avalanche photo detectors (SASPDs). Optimal coupling between the integrated device and the fiber arrays is achieved with two six-axis alignment stages.
\\
\\[\baselineskip]
\textbf{Chip fabrication.} The waveguides were written inside high purity fused silica (Corning 7980ArF grade) using a RegA 9000 seeded by a Mira Ti:Sapphire femtosecond laser. The pulse duration was $150 fs$ at $800 nm$ with a repetition rate of $100 kHz$ and a pulse energy of $200 nJ$. The pulses were focused $370 \mu m$ under the sample surface using a $NA=0.6$ objective while the probe was translated with a constant speed of $6 cm/min$ by high-precision bearing stages (ALS130, Aerotech Inc.). The mode field diameter of the guided modes is $22 \mu m \times 22 \mu m$ at $789 nm$. Propagation loss was measured at $0.3 dBcm^{-1}$ and birefringence was measured in the order of $B=10^{-7}$. Measured coupling with the input fibers ($850 nm$ PM-fibers, OZ Optics) was $-6.7 dB$ while losses at the output facet (graded-index multimode with $50 \mu m$ core diameter) were negligible.
\\[\baselineskip]
\textbf{Unitary matrix reconstruction.} To verify the experimentally obtained output distributions with theoretical data, the random $5 \times 5$ unitary matrices had to be reconstruced. Extracting the actual unitary transformation experimentally~\cite{Peruzzo2011,Thomas-Peter2011,Laing2012} is important as small imperfections in the fabrication process can lead to significant changes in the transformation matrix. Since the position of beam splitters and phase shifters were known for our integrated devices and the output loss was experimentally extracted, the fitting of the 19 unknown variables (eight beam splitter ratios and eleven phase shifts) could be done via single-photon data and two-photon HOM-dips~\cite{Laing2012}. All errors are conservatively computed by assuming Poissonian error propagation.

\begin{acknowledgments}
The authors are grateful to S. Aaronson, {\v C}. Brukner and M. Ringbauer for discussions. We acknowledge support from the European Commission, Q-ESSENCE (No 248095), QuILMI (No 295293) and the ERA-Net CHIST-ERA project QUASAR, the German Ministry of Education and Research (Center for Innovation Competence program, grant 03Z1HN31), the John Templeton Foundation, the Austrian Nano-initiative NAP PlatonAustrian Science Fund (FWF): (SFB-FOCUS) and (Y585-N20) and the doctoral programme CoQuS, and the Air Force Office of Scientific Research, Air Force Material Command, United States Air Force, under grant number FA8655-11-1-3004.\end{acknowledgments}

\end{document}